\documentclass[aps,pra,floatfix,reprint]{revtex4-1}
\usepackage{amsmath}
\usepackage{amsfonts}
\usepackage{amssymb}
\usepackage{graphicx}
\usepackage{color}
\usepackage{placeins}
\usepackage[colorlinks]{hyperref}
\usepackage{multirow}
\usepackage[caption=false]{subfig}
\usepackage[capitalize]{cleveref}

\usepackage{yypreamble}

\hypersetup{urlcolor=blue,linkcolor=black}

\renewcommand{\avg}[1]{\langle{#1}\rangle}

\newcommand{\YYrev}[1]{}

\begin{document}

\title{Shelving-style QND phonon-number detection in quantum optomechanics}

\author{Yariv Yanay}
\author{Aashish A. Clerk}
\affiliation{Department of Physics, McGill University, Montreal, Canada H3A 2T8}

\date{\today}

\begin{abstract}
We propose a new method for optomechanical quantum non-demolition (QND) detection of phonon number, based on a ``shelving" style measurement. The scheme uses a two-mode optomechanical system where the frequency splitting of the two photonic modes is near-resonant with the mechanical frequency. The combination of a strong optical drive and the underlying nonlinear optomechanical interaction gives rise to spin-like dynamics which facilitate the measurement. This approach allows phonon number measurement to be accomplished  
parametrically faster than in other schemes which are restricted to weak driving.  The ultimate power of the scheme is controlled by the size 
of the single photon optomechanical cooperativity.
\end{abstract}

\maketitle


\section{Introduction}
Quantum optomechanical systems, where photons interact with mechanical motion, have been the recent subject of intense experimental and theoretical activity \cite{Aspelmeyer2014a}.  They have enabled the exploration of a wide range of effects, including the cooling of macroscopical mechanical modes to near their quantum ground states \cite{Teufel2011,Chan2011,Peterson2016} and the generation of squeezed states of light \cite{Brooks2012,Safavi-Naeini2013, Purdy2013} and mechanical motion \cite{Wollman2015,Pirkkalainen2015,Lecocq2015}.

\begin{figure}[!h] 
   \centering
   	\subfloat{\includegraphics[width=\columnwidth]{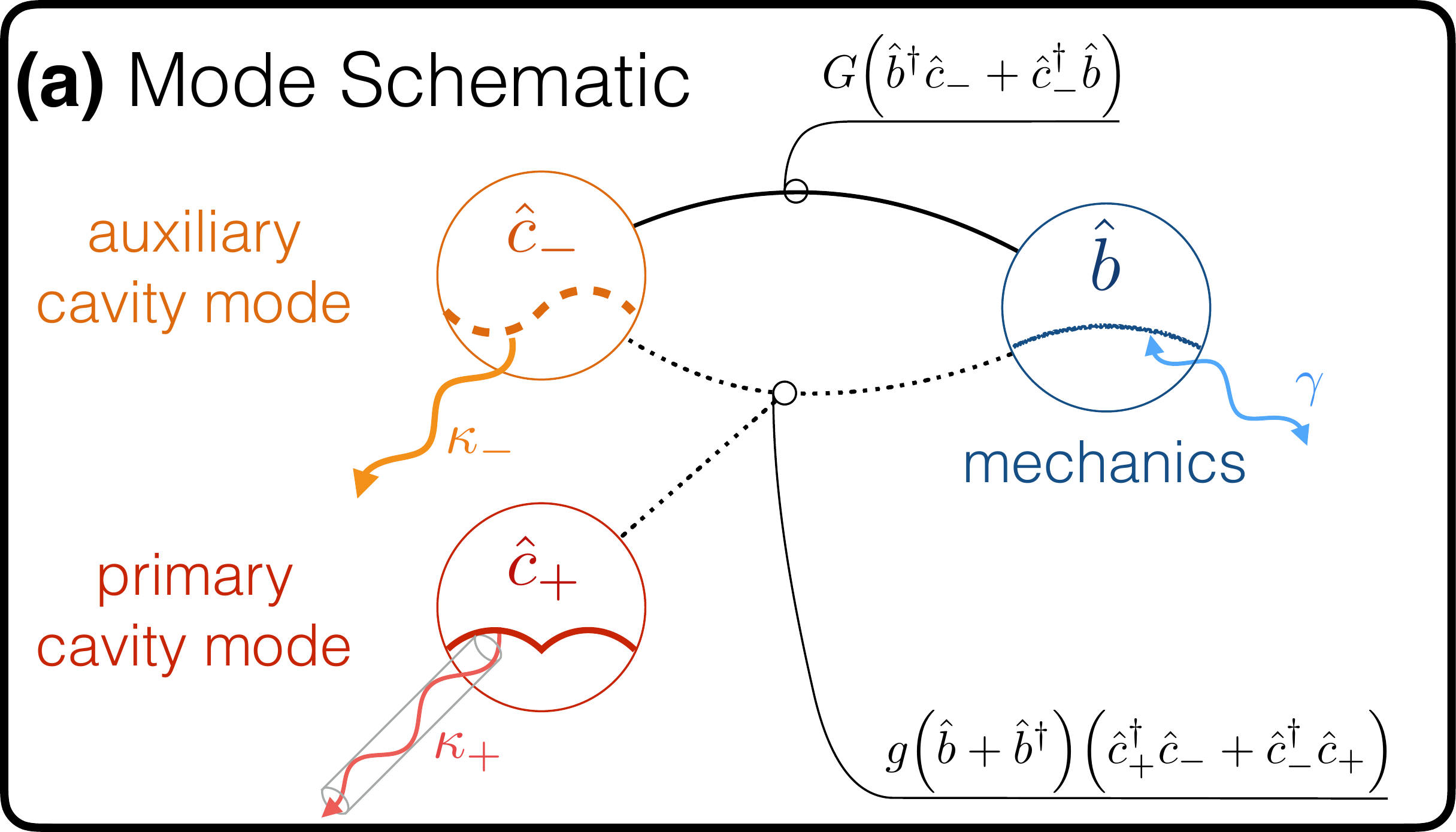}\label{fig:modeschematic}}
   
   	\subfloat{\includegraphics[width=\columnwidth]{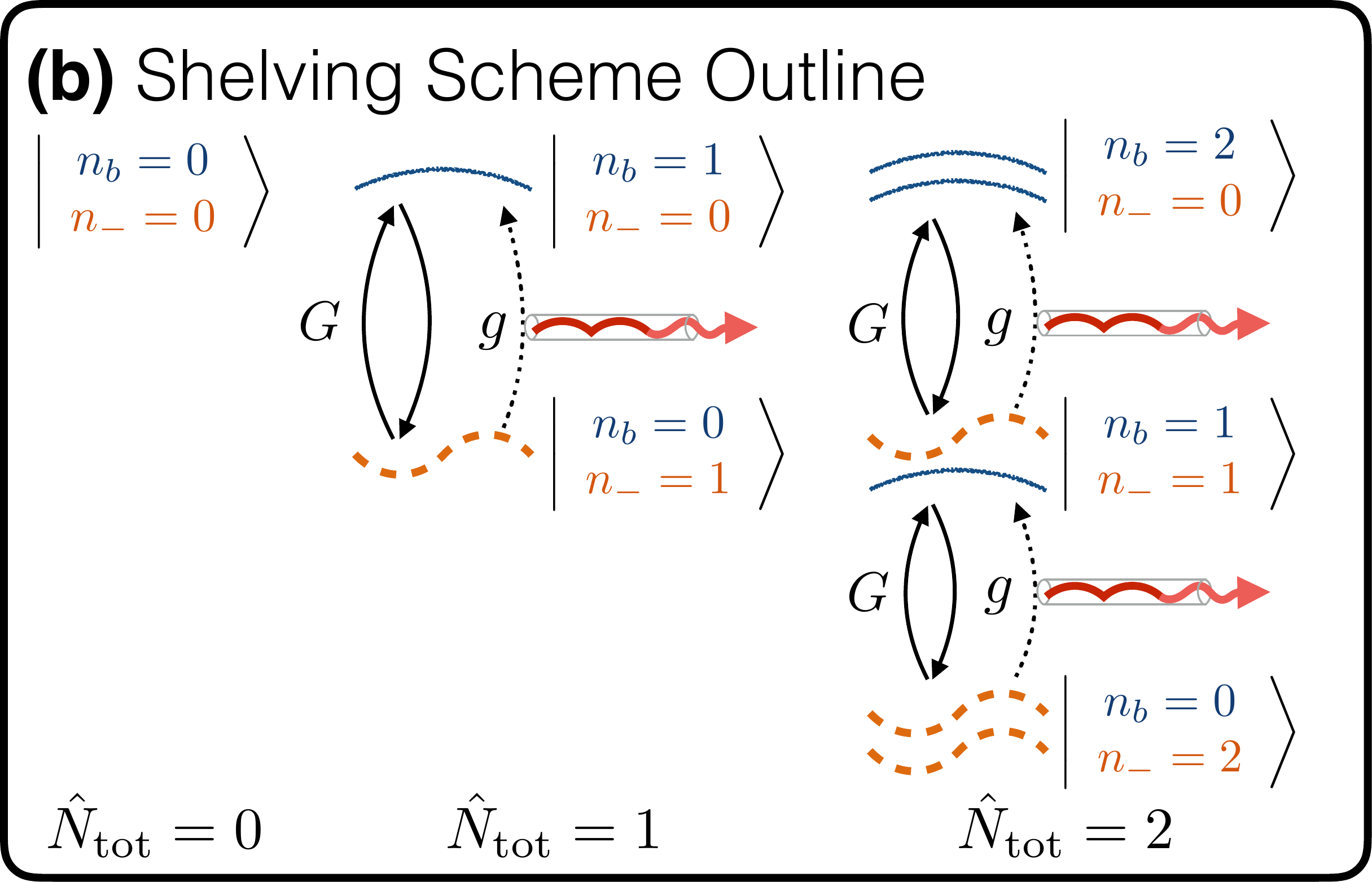}\label{fig:shelvingscheme}}
   
   \caption{{\bf (a)} Schematic of the driven two-cavity optomechanical system described by \crefrange{eq:Htot}{eq:Hint}.  Two photonic modes
   ($\hat c_{\pm}$) are coupled nonlinearly to a mechanical mode $\hat{b}$ (amplitude $g$).  In addition, a strong drive on the 
   primary $\hat{c}_+$ mode
   realizes a beam-splitter interaction between $\hat{b}$ and $\hat c_{-}$ (amplitude $G$).  
       $\gk_{\pm}$ denote the photonic mode damping rates, $\gamma$ the mechanical damping rate.
      {\bf (b)} Schematic of measurement scheme. $G$ causes excitations to oscillate between the mechanics and the auxiliary
       $\hat{c}_-$ mode.  The nonlinear interaction converts a  $\hat{c}_{-}$ photon into a phonon and a $\hat{c}_{+}$ photon; this photon
       is then detected.  The scheme allows a QND measurement of $\hat N_{\rm tot} = \hat{c}_-^\dagger \hat{c}_- + \hat{b}^\dagger \hat{b}$,
       and thus  QND measurement of the initial phonon number.  
}
   \label{fig:diagram}
\end{figure}

 A key unrealized goal in optomechanics is the quantum non-demolition (QND)
measurement of mechanical phonon number, something that would directly reveal the energy quantization of the mechanics \cite{Santamore2004,Martin2007}.  The most promising proposals involve optomechanical systems where a single mechanical resonator couples to two photonic modes, as depicted schematically in Fig.~\ref{fig:diagram}a.  Such systems can in principle be used for phonon-number detection
both in the regime where the mechanical frequency is much smaller than the splitting between the optical modes 
\cite{Thompson2008,Jayich2008,Miao2009,Gangat2011,Yanay2016a}, 
and in the regime where it is comparable to this splitting \cite{Ludwig2012,Komar2013,Basiri-Esfahani2012}.  Two-cavity
optomechanical systems have been realized experimentally using a variety of platforms \cite{Thompson2008,Paraiso2015,Toth2016}.

In this work, we revisit the two-mode system of Fig.~\ref{fig:modeschematic}, and focus on the regime where the mechanics are almost resonant with the frequency difference of the photonic modes, and where one photonic mode is strongly driven.  This strong drive generates a coherent, many-photon optomechanical interaction which cannot be treated perturbatively. We show that this regime allows a new approach to phonon number measurement, one which offers advantages over previous proposals (namely a much faster measurement).
The regime we study differs from that of Ludwig et al.~\cite{Ludwig2012}, who considered more non-resonant interactions and weak driving, such that a pertubative treatment was possible. It also contrasts from the work of K\'om\'ar et al.~\cite{Komar2013}, who considered a perfectly resonant system and weak optical driving, but did not consider phonon number measurement physics.

 Our measurement scheme is sketched in \cref{fig:shelvingscheme}, and is conceptually similar to the highly successful electron shelving technique used in trapped-ion systems \cite{Dehmelt1968,Nagourney1986,Javanainen1986}, as explained in \cref{sec:shelvinganalogy}. The measurement starts by turning on the strong optical drive on the ``primary" mode. This induces coherent oscillations between phonons and photons in the undriven ``auxiliary" mode, oscillations which necessarily conserve the total number $\hat{N}_{\rm tot}$ of aux-mode and mechanical excitations. The remaining nonlinear interaction then allows one to use the output field of the driven optical mode to measure $\hat{N}_{\rm tot}$ without changing its value - a true QND measurement.  This is then effectively a measurement of mechanical phonon number, as without spurious dissipative effects,  $\hat{N}_{\rm tot}$ is simply equal to the initial phonon number.  

As with other schemes, our approach still requires a relatively strong single-photon optomechanical coupling (i.e. it must be large compared to the geometric mean of the two photonic damping rates) to resolve mechanical Fock states. However, it allows measurement rates that are parametrically larger than the perturbative regime considered in Ref.~\onlinecite{Ludwig2012}, or in the adiabatic regime of a position-squared measurement considered in Ref.~\onlinecite{Thompson2008,Jayich2008,Miao2009}.
This faster rate could be of significant utility in new experimental designs of two-cavity optomechanical systems where one photonic mode is explicitly engineered to have extremely low dissipation \cite{Paraiso2015, Toth2016}.  In such systems, the ultimate limit to Fock state detection comes from the requirement that the measurement rate be faster than the heating rate of the mechanics due to their intrinsic dissipation.  As we show, the relevant parameter controlling the utility of our scheme in this limit is the single-photon cooperativity \cite{Aspelmeyer2014}, \YYrev{though this regime is still experimentally inaccessible.}

The remainder of the paper is arranged as follows. In \Cref{sec:model} we introduce the theoretical model of our two-mode optomechanical system, and discuss the parameter regime of interest.  In \Cref{sec:pseudospin} we introduce a mapping 
to an effective spin system that provides a convenient way to describe the system dynamics; we focus here on the ideal case
where the only dissipation is due to the coupling to the input/output port used for the measurement.  In \Cref{sec:meas} we discuss the measurement protocol.  Finally, in \Cref{sec:gamma} we study the effects of the unwanted dissipation both in the aux mode and in the mechanics, and the limits that they impose on QND measurement.

\section{Model}
\label{sec:model}

The optomechanical system of interest (c.f.~Fig.~\ref{fig:diagram}a) consists of two photonic modes and a single mechanical mode.  Its coherent dynamics is described by the Hamiltonian
\begin{equation}
	\begin{split}
	\hat  H_{S} & = \hat  H_{\rm opt} + \hat  H_{\rm m} + \hat{H}_{\rm int}.
	\label{eq:Htot}
	\end{split}
\end{equation}
The first term describes the two photonc eigenmodes in the absence of optomechanical coupling, and is ($\hbar = 1$ throughout)
\begin{equation}\begin{split} 
	\hat  H_{\rm opt} 
		 & = \left( \omega_{c}-J \right) \hat a_{+}\dg\hat a_{+} + \p{\omega_{c} + J}\hat a_{-}\dg\hat a_{-},
		 \label{eq:Hopt}
\end{split}\end{equation}
where $\hat a_{\pm}$ are annihilation operators for the two photonic modes (frequencies $\omega_c-J$ and $\omega_c + J$ respectively).  
We call these the primary ($\hat a_{+}$) and auxiliary ($\hat a_{-}$) modes.  

The second term in Eq.~(\ref{eq:Htot}) describes a mechanical mode with frequency 
$\omega_{\rm m}$ and annihilation operator $\hat b$:
\begin{equation}
\hat H_{\rm m} = \omega_{\rm m}\hat b\dg\hat b = \p{2J - \gd\gO}\hat b\dg\hat b.
\end{equation}
We have introduced the parameter $\gd\gO = 2J - \omega_m$ which represents the mismatch between the photonic mode splitting and the mechanical frequency.  Similar to reference \cite{Ludwig2012}, we will be interested
in the regime where the mechanical frequency is similar in magnitude to $2J$, and hence $| \gd \gO | \ll J$.  

Finally, the last term in \cref{eq:Htot}  describes the optomechanical coupling, which takes the form of a mechanically-mediated tunneling interaction between the two modes: 
\begin{equation}\begin{split}
\hat{H}_{\rm int} & = -g\p{\hat b + \hat b\dg}\p{\hat a_{-}\dg\hat a_{+} + \hat a_{+}\dg\hat a_{-}},
\label{eq:Hint}
\end{split}\end{equation}
where $g$ is the single-photon optomechanical coupling amplitude.
For the specific case of a membrane-in-the-middle style optomechanical system \cite{Thompson2008}, a detailed derivation of this model is given in Refs.~\onlinecite{Jayich2008,Cheung2011};  its limitations are discussed in Refs.~\onlinecite{Cheung2011,Xuereb2012}.

We next use a standard treatment to describe the effects of dissipation on the system \cite{Gardiner2004,Clerk2010}. We assume that each optical mode is coupled to an independent zero-temperature Markovian reservoir, giving rise to an amplitude damping rate of $\gk_{\pm}/2$ for $\hat a_{\pm}$, respectively.  The mechanics are similarly coupled to a thermal Markovian reservoir characterized by a thermal occupancy $n_{\rm th}$, leading to an amplitude damping rate of $\gamma/2$ for $\hat b$.  Note that for a membrane-in-the-middle type system, treating the optical modes as seeing independent reservoirs can be problematic \cite{Yanay2016a}, as it neglects a possible dissipation-induced coupling of the optical normal modes.  Such a coupling will not play a role in our scheme, as it is suppressed by the optical mode splitting $2J$, while the main optomechanical coupling is resonant.

Our measurement scheme will involve applying a strong coherent drive to the primary, $\hat{a}_+$, mode near its resonance frequency $\omega_c - J$.  For simplicity, we assume the case of a perfectly resonant drive, as qualitatively similar results are obtained for a detuned drive. In this case, it is useful to move to a rotating frame, and also make a standard displacement transformation on the optical modes:
\begin{equation}\begin{gathered}
\hat a_{-} \to e^{-i\p{\omega_{c} + J - \gd\gO/2}t}\hat c_{-}, 
 	\qquad \hat b \to e^{-i\p{\omega_{\rm m} + \gd\gO/2}t}\hat b,
\\ \hat a_{+} \to e^{-i\p{\omega_{c} - J}t}\p{\avg{\hat a_{+}}_{g=0} + \hat c_{+}}, \quad G = g\avg{\hat a_{+}}_{g=0}.
\end{gathered}
\label{eq:IntPic}
\end{equation}
Here, $\avg{\hat a_{+}}_{g=0}$ is average primary-mode amplitude induced by the drive in the absence of optomechanical coupling, and
$\hat{c}_+$ is the displaced annihilation operator for this mode in the rotating frame. 

In this frame we find $\hat H_{S} = \hat H_{\rm eff} + \hat H_{\rm NR}$ where
\begin{align}
\begin{split}\hat H_{\rm eff} & =  \half[\gd\gO]\mat{\hat c_{-}\dg\hat c_{-} - \hat b\dg\hat b}  -G\mat{\hat c_{-}\dg\hat b + \hat b\dg\hat c_{-} }
	\\ & \quad -g\mat{  \hat b\dg\hat c_{+}\dg\hat c_{-}  +  \hat c_{-}\dg \hat c_{+}\hat b  }\end{split}
\label{eq:Heff}
\\ \hat H_{\rm{NR}} & =  -e^{i \left( 4 J - \gd\gO \right) t}\hat b\dg\hat c_{-}\dg \mat{G + g\hat c_{+}} +\hc
\label{eq:HNR}
\end{align}
Finally, we assume that the optical mode gap $2J$ is sufficiently large that we can safely make a rotating wave approximation and drop the non-resonant interactions described by $\hat H_{\rm{NR}}$.  This requires in practice $J\gg |\gd\gO|, G, g,\gk_{+}$.  We are left
with the interactions depicted in Fig.~\ref{fig:diagram}a:  a beam splitter interaction between the mechanics and the auxiliary mode, and a nonlinear interaction involving all three modes.

Note that in the absence of driving, i.e. $G=0$, the photonic modes are only driven by vacuum noise, and the nonlinear optomechanical interaction in \cref{eq:Heff} effectively vanishes.  Therefore, without driving the mechanics are completely decoupled from the photonic modes.  We imagine that before the measurement is turned on by driving the primary photonic mode, the mechanics are first prepared close to the ground state, e.g. via standard cavity cooling techniques.  The scheme we describe in what follows then allows one to measure the mechanical phonon number. \YYrev{This is a weak continuous QND measurement, and we analyze it in the standard way \cite{Walls1997}. }

\section{Dynamics and effective spin description\label{sec:pseudospin}}

\subsection{Conserved excitation number}

The nonlinear interaction in \cref{eq:Heff} has the form of a standard three-wave mixing Hamiltonian, describing e.g. a non-degenerate parametric amplifier with a dynamical pump mode.  It has three conserved excitation numbers (typically known as Manley-Rowe constants of motion \cite{Manley1956}) reflecting the SU(1) and SU(1,1) symmetries of the interaction.  Including the beam-splitter interaction reduces the symmetry of the Hamiltonian, and $\hat{H}_{\rm eff}$ has only a single conserved excitation number $\hat{N}_{\rm tot}$:
\begin{equation}
	\hat N_{\rm tot} = \hat b\dg\hat b + \hat c_{-}\dg\hat c_{-},
	\hspace{0.6 cm}
	\left[ \hat{N}_{\rm tot}, \hat{H}_{\rm eff} \right] = 0
\end{equation}
Note that as $\hat{N}_{\rm tot}$ is independent of the primary mode, it continues to be conserved even if this mode is damped.

In what follows, we first analyze the ideal case where the only dissipation in the system is due to its coupling to the input-output port used for the measurement.  This corresponds to $\kappa_{+} \neq 0$, while $\gamma = \kappa_{-} \simeq 0$.  Starting with this limit will let us present a clear picture of how our proposed phonon number measurement operates:  the conservation of $\hat{N}_{\rm tot}$ in this limit is at the heart of the shelving scheme depicted in \cref{fig:diagram}b.  In this ideal case, we find it is always possible to determine the initial mechanical phonon number by simply measuring for long enough; this is akin to standard shelving measurements. 

The regime $\kappa_+ \gg \gamma, \kappa_{-}$ is also relevant experimentally.  While mechanical damping rates $\gamma$ are typically orders of magnitude smaller than photonic damping rates, it might first seem surprising that one could achieve $\kappa_{-} \ll \kappa_{+}$.  However, recent experiments with microwave-circuit optomechanics have achieved precisely this kind of regime through clever experimental design. \YYrev{Current designs have reached $\gk_{+}\approx 40 \gk_{-}$ \cite{Toth2016}, though the spurious dissipation is still strong compared with the optomechanical coupling $g$.} 

We analyze the effects of non-zero $\gamma, \gk_{-}$ in \cref{sec:meas}. As expected, they limit the maximum possible measurement time, putting constraints on whether phonon number detection is achievable.

\subsection{Mapping to an effective spin system}

To understand the dynamics of our system, it is helpful to make the conservation of $\hat N_{\rm tot}$ explicit by representing the mechanical and aux-modes by an effective spin (a Schwinger boson representation in reverse).  We thus introduce the effective (dimensionless) angular momentum operators $\vec{\hat{J}}$ via
\begin{equation}\begin{gathered}
\hat J_{+} = \hat b\dg\hat c_{-} \qquad \hat J_{-} = \hat c_{-}\dg\hat b
\\ \hat J_{\pm} \equiv \hat J_{x}\pm i\hat J_{y} \qquad \hat J_{z} = \half\p{\hat b\dg\hat b - \hat c_{-}\dg\hat c_{-}}.
\end{gathered}\end{equation}
These operators satisfy standard angular momentum commutation relations.  A direct calculation shows:
\begin{equation}
	\hat{J^2} =  \frac{\hat{N}_{\rm tot}}{2}  \left( \frac{\hat{N}_{\rm tot}}{2}+1 \right).
	\label{eq:SpinSize}
\end{equation}
The conservation of $\hat{N}_{\rm tot}$ now just corresponds to our effective spin having a fixed length.

With these operators, the effective Hamiltonian is
\begin{equation}
	\begin{split}
	\hat  H_{\rm eff} & = - \vec B \cdot \vec{\hat J} - g\mat{\hat J_{-}\hat c_{+} + \hat c_{+}\dg\hat J_{+}}.
	\label{eq:Hspin}
	\end{split}
\end{equation}
where the effective magnetic field is
\begin{equation}
\vec B = 2G\mathbf{e}_{x} + \gd\gO\mathbf{e}_{z} \equiv B\mathbf{e}_{B}
\end{equation}
with $\mathbf{e}_{x},\mathbf{e}_{y},\mathbf{e}_{z}$ the unit vectors in the respective directions.
As the $\hat{c}_+$ mode is damped, we see that $\hat{H}_{\rm eff}$ takes the form of a generalized spin-boson model.
The quadratic terms in the original version of $\hat{H}_{\rm eff}$ now correspond to a Zeeman field on our spin.  The order $g$ nonlinear term in 
$\hat{H}_{\rm eff}$ now corresponds to a system-bath coupling where a bath boson can be created while adding angular momentum in the $z$ direction to the spin.

The Heisenberg-Langevin equations of motion take the form
\begin{align}
\dot{\hat c}_{+} = -\half[\gk_{+}]\hat c_{+} + \sqrt{\gk_{+}}\hat c_{\rm in} + ig\hat J_{+}
\label{eq:eomspinc}
\\ \vec{\dot{\hat J}} =  \vec{\hat J}\times \p{\vec B + g\br{\mathbf{e}_{-}\hat c_{+} + \mathbf{e}_{+}\hat c_{+}\dg}}
\label{eq:eomspinJ}
\end{align}
where $\mathbf{e}_{\pm} = \mathbf{e}_{x}\pm i \mathbf{e}_{y}$. The operator $\hat c_{\rm in}$ describes the vacuum fluctuations entering the system from the 
input-output line coupled to the principal $+$ mode (i.e.~from the input-output waveguide that will be used for measurement); it has
zero mean and correlation functions
\begin{equation}
\avg{\hat c_{\rm in}\dg\p{t}\hat c_{\rm in}\p{t\pr}} = 0 \qquad \avg{\hat c_{\rm in}\p{t}\hat c_{\rm in}\dg\p{t\pr}} = \gd\p{t-t\pr}.
\end{equation}

When the optomechanical drive is initially turned on, $\hat{N}_{\rm tot} = \hat{n}_b$, where $\hat{n}_b \equiv \hat{b}^\dagger \hat{b}$ is the initial mechanical phonon number.  From Eq.~(\ref{eq:SpinSize}), we see that the phonon number manifests itself in the dynamics by directly setting the size of the effective spin $\vec{\hat{J}}$.  
The nonlinear interaction creates photons in the principal mode $\hat{c}_+$ in a way that depends on $\hat J_{\pm}$, and hence on the magnitude of
$\vec{\hat{J}}$. Thus, by monitoring the output field from this mode, one can determine the initial phonon number.  We stress that the ``backaction" from the this measurement only alters the {\it direction} our effective spin points in, but not its size.  The measurement is thus QND.

For easy reference, a synopsis of the mapping from bosons to spins is provided in \cref{tab:spincomp}.

\begin{table}[t]
\caption{\label{tab:spincomp}Summary of the spin formalism}
\renewcommand{\arraystretch}{2}
\begin{ruledtabular}
\begin{tabular}{ccp{5.5cm}}
	Spin term & \multicolumn{2}{c}{Optomechanical Equivalent}
		\\ \hline
	$\hat J_{z}$ & $\half[\hat b\dg\hat b - \hat c_{-}\dg\hat c_{-}]$ &  
		\multirow{2}{5.5cm}{The $z$-projection of the effective spin represents the difference
		between the number of phonons and number of aux-mode quanta.}
\\ $\hat J_{+}$, $\hat J_{-}$ & $\hat b\dg\hat c_{-}$, $\hat c_{-}\dg\hat b$
\\ $B_{x}$ & $2G$ & The many-photon optomechanical interaction $G$ exchanges photons and phonons, 
	and hence acts like a magnetic field in $\mathbf{e}_{x}$ direction.
\\ $B_{z}$ & $\gd\gO$ &The mismatch between the photonic mode splitting and the mechanical frequency acts as an $\mathbf{e}_{z}$ field.
\end{tabular}
\end{ruledtabular}
\end{table}

\subsection{Dynamics in the limit $g\ll \kappa_{+}$\label{sec:adiabdyn}}

We consider the system dynamics in the typical to experiment, ${\gk_{+}\gg g}$. We begin by solving \cref{eq:eomspinc} to find
\begin{equation}\begin{split}
\hat c_{+}\p{t} & = e^{-\half\gk_{+}t}\hat c\p{0} + \hat \zeta\p{t}
\\ & \quad + ig \int_{0}^{t}{d\tau}\;e^{-\half \gk_{+}\tau}\hat J_{+}\p{t-\tau}.
\label{eq:ct}
\end{split}\end{equation}
where  $\hat \zeta\p{t} = \sqrt{\gk}\int_{0}^{t}{d\tau} \;e^{-\half \gk_{+}\p{t-\tau}}\hat c_{\rm in}\p{\tau}$.

We see that $\hat{c}_+(t)$  depends on the behaviour of $\vec{\hat J}$ at earlier times, and that the range of this memory effect is $\gk_{+}^{-1}$. 
We can solve for the spin dynamics during this short interval to a good approximation by ignoring the effects of $g$, as
\begin{equation}
	\vec{\hat J}\p{t-\tau} = e^{i\vec B\cdot\vec{\hat J}\tau}\p{\vec{\hat J}\p{t} + O\p{g\tau}}e^{-i\vec B\cdot\vec{\hat J}\tau}.
\end{equation}
As long as $\tau$ is at most $\sim \kappa_+$, we can drop the $O\p{g\tau}$ term as it is proportional to the small parameter
\begin{equation}
	\eta = 2g/\gk_{+}.
\end{equation}

With this approximation, the solution for $\hat{c}_+(t)$ takes the simpler form:
\begin{equation}
	\hat c_{+}\p{t} = e^{-\half\gk_{+}t}\hat C_{0} + \hat \zeta\p{t} + i\eta \hat J_{c}\p{t} + O\p{\eta}^{2}.
\label{eq:ctchi}
\end{equation}
Here, the non-Hermitian operator $\hat{J}_c$ is linear in the components of $\vec{\hat J}$, and given by
\begin{equation}\begin{gathered}
	\hat J_{c} =  
	\br{\frac{B_{x} \vec B + \half[\gk_{+}] \mathbf{e}_{+}\times \vec B + \p{\half[\gk_{+}]}^{2}\mathbf{e}_{+}}{\p{\half[\gk_{+}]}^{2} + B^{2}}}\cdot \vec{\hat J}.
\label{eq:chi}
\end{gathered}\end{equation}
The initial transient behaviour of $\hat{c}_+$ is described by 
${\hat C_{0} = \hat c_{+}\p{0} - i \eta \hat J_{c}\p{0}}$.

\Cref{eq:ctchi} implies that the primary mode (and its output optical field) will be sensitive to components of $\vec{\hat J}$ in a manner that depends on both $\vec{B}$ and $\kappa_{+}$.  In the extreme Markovian limit ${\gk_{+}\gg B}$ we find $\hat J_{c} \to \hat J_{+}$, implying that the $\hat{c}_+$  is able to measure both transverse components of our effective spin.  In contrast, in the limit $B\gg\gk_{+}$, the $\hat{c}_+$ mode is only able to respond to the non-rotating component of the spin, and hence $\hat J_{c} \to (B_x / B) \mathbf{e}_{B}\cdot\hat J$.

It is also instructive to consider the dynamics in terms of the reduced density matrix describing the spin. 
The interaction picture density matrix is defined in terms of the Schr\"odinger-picture density matrix 
$\hat \rho_{J} = \Tr_{\hat c_{+}}\br{\hat\rho}$ via
\begin{equation}
	\tilde \rho_{J}(t) = e^{-i\vec B\cdot {\vec{\hat{ J}}}t}\hat \rho_{J}(t)e^{i\vec B\cdot {\vec{\hat{ J}}}t}.
\end{equation}
We again consider
the $g \ll \kappa_+$ limit, and derive a weak-coupling master equation, retaining terms to leading order in $\eta$. 
For simplicity, we also consider the typical limit where $B \gg \p{\eta g = 2g^2 / \kappa_+}$, allowing us to drop rapidly oscillating terms (i.e.~make a secular approximation). 

We use $\hat J_{\parallel} = \mathbf{e}_{B}\cdot \vec{\hat J}$ to denote the projection of the spin along the axis of the effective magnetic field $\vec{B}$
, and use $\hat J_{\pm}\pr$ to denote raising and lowering operators for $\hat J_{\parallel}$.
Defining $\mathcal D\br{\hat a}\cdot{\hat \rho} = \hat a\hat \rho \hat a\dg - \half\p{\hat a\dg\hat a \hat\rho + \hat\rho \hat a\dg\hat a}$ as the 
standard Lindblad superoperator, 
we find
\begin{equation}\begin{split}
	\dot{\tilde \rho}_{J} & = i \br{\tilde \rho_{J}, \hat H_{1}} +  
	\\ &\left( 
		\Gamma_{\varphi} \mathcal D\br{\hat J_{\parallel}} 
	  + \Gamma_{+} \mathcal D\br{\hat J_{+}\pr}  
	+ \Gamma_{-} \mathcal D\br{\hat J_{-}\pr} 
	\right) \cdot{\tilde \rho_{J}}.
\label{eq:Jmaster}
\end{split}\end{equation}
This has the form of a standard master equation for a spin coupled to a bath.  The first term describes a modification of the coherent dynamics, 
described by
\begin{equation}
	 \hat H_{1} = \tfrac{g^{2}B}{B^{2}+ \p{\half[\gk_{+}]}^{2}}\br{\tfrac{B_{z}}{B}\hat J_{\parallel} - 
	 \tfrac{B^{2} + B_{z}^{2}}{2B^{2}}\p{\vec{\hat J}^{2} - \hat J_{\parallel}^{2}}}.
\end{equation}
The remaining terms describe dephasing (rate $\Gamma_{\varphi}$),
and spin raising / lowering transitions along the axis $\mathbf{e}_{B}$ (rates $\Gamma_{\pm}$).  The dissipative rates are

\begin{equation}\begin{gathered}
\Gamma_{\varphi} =  \frac{B_{x}^{2}}{B^{2}}\frac{4g^{2}}{\gk_{+}},
\qquad \Gamma_{\pm} = \frac{\left(B \pm B_{z}\right)^{2}}{2B^{2}} \frac{g^{2}\gk_{+}}{B^{2} + \p{\half[\gk_{+}]}^{2}}.
\label{eq:effrates}
\end{gathered}\end{equation}

Note that unless $B_z = B$, the vacuum noise driving the $\hat{a}_+$ mode is able to cause spin flips upwards in energy:  this is possible because we are strongly driving the $\hat{a}_+$ mode, and can be viewed as being analogous to the phenomenon
of  ``quantum heating" \cite{Dykman2011,Peano2010,Lemonde2015}.  Note also that we always assume that the drive frequency $\omega_{c} - J$ is much larger than $G, |\gd\gO|$.  

 From \cref{eq:Jmaster} we find that the steady state of the spin is simply a thermal state, 
\begin{equation}
\tilde \rho_{J} \to e^{B\hat J_{\parallel}/T_{\rm eff}}/\Tr\br{e^{B\hat J_{\parallel}/T_{\rm eff}}}
\label{eq:rhoJss}
\end{equation}
with an effective temperature
 \begin{equation}
B/T_{\rm eff} = 2 \ln\p{\frac{B + B_{z}}{B - B_{z}}}.
\label{eq:Teff}
 \end{equation}
 Again, unless $B_z = B$, the steady state will not correspond to the spin being in its ground state (due to the previously mentioned quantum heating physics).  Note that in the case where the effective field is completely in the $x$ direction (i.e.~the detuning $\gd\gO$ vanishes), our master equation describes an effective infinite temperature, and the spin will be completely depolarized in the steady state. \YYrev{This corresponds to a mixed state with equal probability for all Fock states adding up to the conserved $\hat N_{\rm tot}$.}
  
\begin{figure}[t] 
   \centering
   \includegraphics[width=\columnwidth]{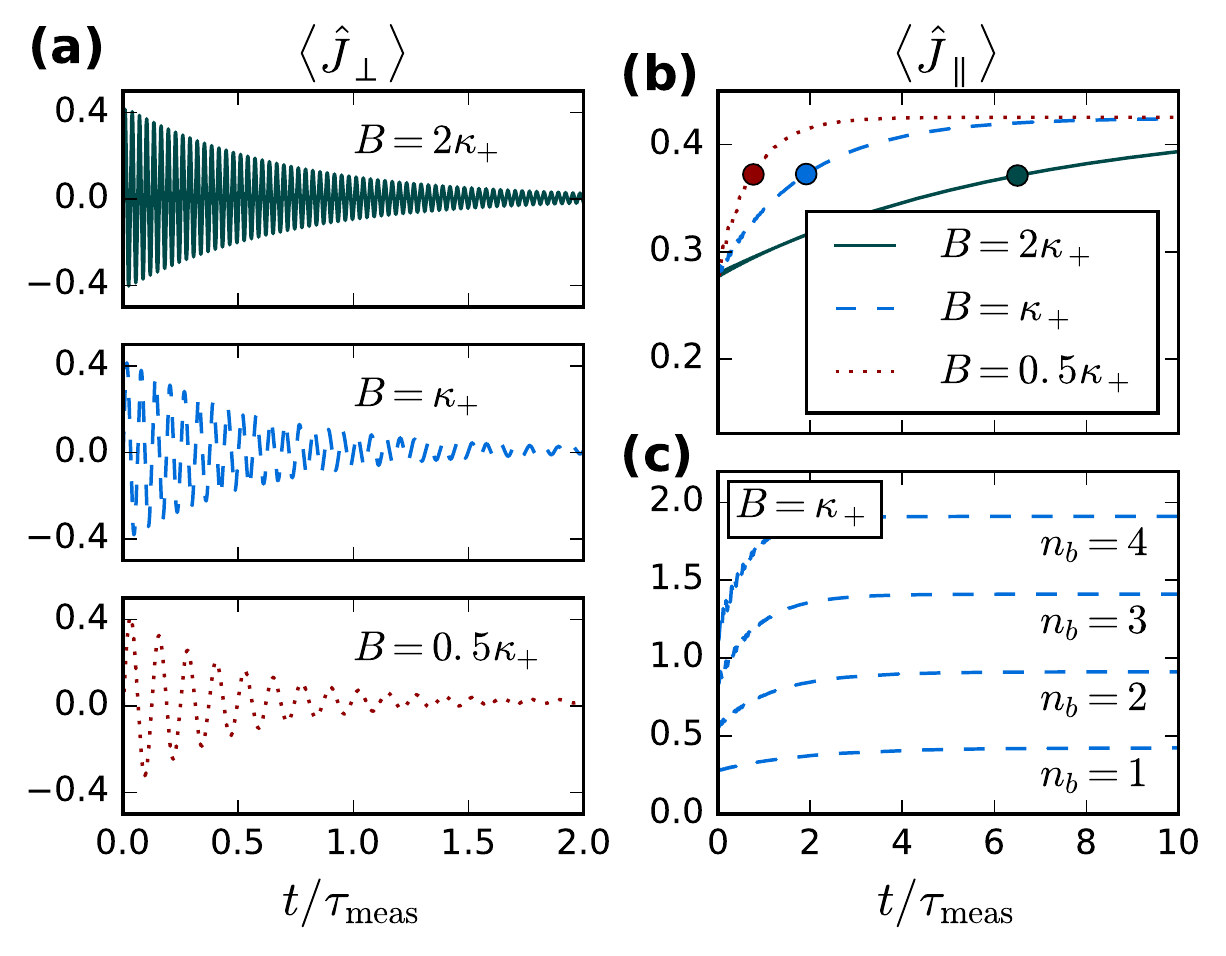} 
   \caption{Behavior of the average of the effective spin, $\avg{\vec{\hat J}}$, 
   as a function of time, for the ideal case where $\gamma = \kappa_{-} = 0$.
   The dynamics described by \crefrange{eq:Jmaster}{eq:effrates} corresponds to standard Markovian dissipative behaviour of a spin
   in an applied field.  
   Here, $\tau_{\rm meas} = \gk_{+}/g^{2}$, and for all plots  $g = 0.1\gk_{+}$ and $G/\gd\gO = 0.75$. For (a)-(b), $n_{b} = 1$ and $B = \sqrt{4G^{2} + \gd\gO^{2}}$ varies, while for (c) $B = \gk_{+}$ and $n_{b}$ varies.  
   {\bf (a)} Dephasing, i.e. decay of transverse $\hat J_{\perp} = \mathbf{e}_{y}\cdot \vec{\hat J}$ component of the spin.  The dephasing rate 
   $\Gamma_{\varphi}^{-1}$ is largely independent of $B$ (c.f.~Eq.~(\ref{eq:effrates})).
   {\bf (b-c)} Relaxation of the parallel spin component $\hat J_\parallel = \mathbf{e}_{B}\cdot\vec{\hat J}$.
   While the relaxation rate is sensitive to $B$, the steady state value of $\hat J_\parallel$ is not, but is instead largely determined
   by the size of the spin (and hence the initial phonon number).  Bold dots in (b) mark the point $\half\p{\Gamma_{+} + \Gamma_{-}} t = 1$
   (c.f.~Eq.~(\ref{eq:effrates})). }
   \label{fig:Jbehave}
\end{figure}

\Cref{fig:Jbehave} illustrates the dissipative spin dynamics described above (see \cref{sec:numerics} for details on numerics). 

\subsection{QND nature of the measurement}

As $\hat{N}_{\rm tot}$ is a conserved quantity, it is clear that our protocol allows for a QND measurement of $\hat{N}_{\rm tot}$: this quantity commutes with the system Hamiltonian, even when we include the coupling $\kappa_+$ to the input-output port used for the measurement.

Perhaps less obvious is that this also corresponds to a QND measurement of the phonon number in the mechanics.  As already discussed, when the control drive is first turned on, the value of $\hat{N}_{\rm tot}$ corresponds to the initial mechanical phonon number $\hat{n}_b$.  For a true QND measurement of phonon number, one would like the final phonon number to return to this value  after the measurement is complete.  This is easily achieved by turning off the large control drive applied to the primary mode.  Once this drive is off, $G=0$, the nonlinear interaction 
(last term in \cref{eq:Heff}) keeps acting until $\hat{n}_{b} = \hat{N}_{\rm tot}$, i.e.~until all the original phonons are returned to the mechanical resonator. 

\YYrev{
To summarize,
\begin{itemize}
\item We begin with the drive off. In the absence of auxiliary photons, the statistics of $\hat N_{\rm tot}$ are identical to those of $\hat n_{b}$.
\item When the drive is turned on, the cavity performs a weak continuous QND measurement of $\hat N_{tot}$. In a conditional treatment, the system will eventually be localized into a well-defined eigenstate of $\hat N_{\rm tot}$.
\item When the drive is then turned off, the final state of $\hat N_{\rm tot}$ is transferred back into $\hat n_b$.
\end{itemize}
The full protocol then acts as a QND measurement of the phonon number $\hat n_b$.}

\section{Description of the measurement output}
\label{sec:meas}

\subsection{Homodyne measurement and measurement time in the absence of spurious dissipation}

Having discussed its dynamics, we now turn to how information on the spin system (and hence the initial photon number) emerges in the output field leaving the primary phononic mode $\hat{c}_+$.  \cref{eq:ctchi} tells us that the output field will be sensitive to particular components of the spin vector $\vec{\hat{J}}$.  
Recall that in our original lab frame, the primary mode is driven at its resonance frequency $\omega_+ = \omega_c - J$.  Information on our effective spin will be optimally encoded in one quadrature of the primary mode output field.  We thus consider a homodyne measurement which directly probes this optimal output field quadrature, again starting with the ideal case where $\gamma = \kappa_{-} = 0$.

From standard input-output theory (and working in the interaction picture defined by Eq.~(\ref{eq:IntPic})), the cavity output field
leaving the $+$ mode is given by
\begin{equation}
	\hat c_{\rm out}(t) = \hat c_{\rm in}(t) - \sqrt{\gk_{+}}\hat c_{+}(t),
\end{equation}
and the phonon number can be obtained from measuring an output quadrature,
\begin{equation}
	\hat X^{\ga}_{\rm out}(t) = \left( e^{i\ga}\hat c_{\rm out}(t) + e^{-i\ga}\hat c_{\rm out}\dg(t) \right) / \sqrt{2},
\label{eq:defX}
\end{equation}
where the angle $\alpha$ parameterizes the choice of quadrature.

\begin{figure}[t] 
   \centering
   \includegraphics[width=\columnwidth]{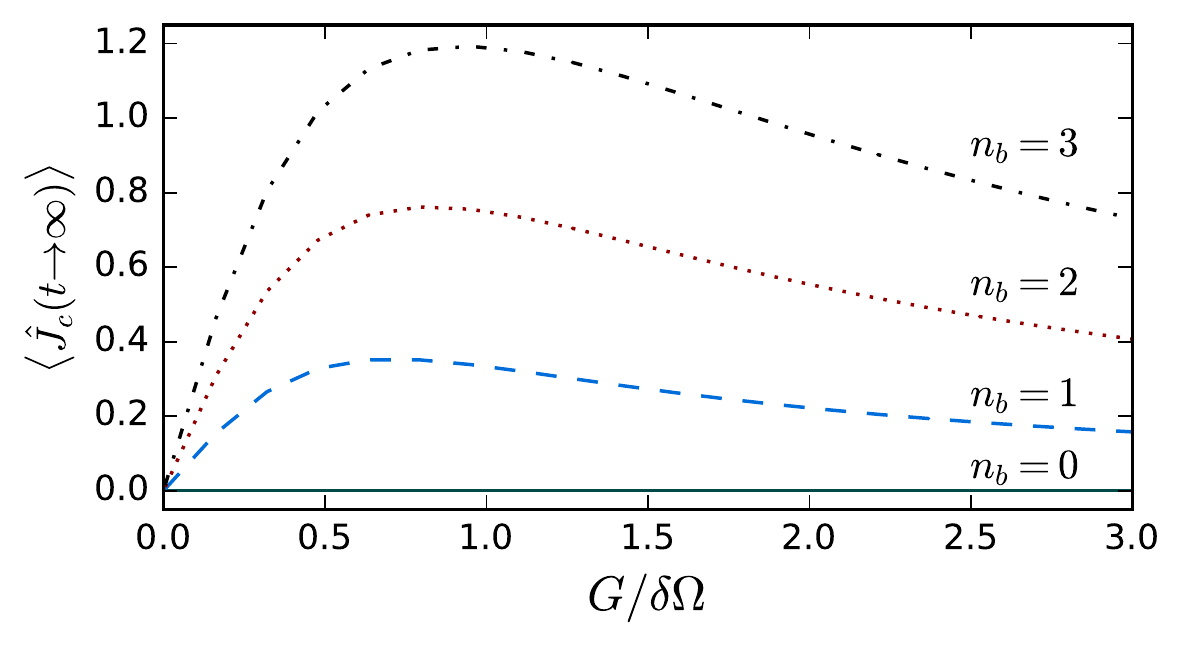} 
   \caption{Steady-state value of $\avg{\hat J_{c}}$ as a function of $G/\gd\gO$. As seen in \cref{eq:Xval}, it is proportional to the amplitude of the measurement output. $G$ is the many-photon optomechanical coupling, and $\gd\gO$ is the mismatch between the mechanical frequency and the photonic mode splitting.  In the spin mapping, $2G /  \gd\gO = B_x / B_z$.   Different curves correspond to different initial phonon numbers $n_b$.  The steady-state value of  $\avg{\hat J_{c}}$ is sensitive to phonon number for a wide range of $G / \gd\gO $.   All curves correspond to the case where there is no spurious dissipation, $\gamma = \kappa_{-}  =0$, and the limit $\sqrt{4G^2 + (\gd \gO)^2} \gg g^2 / \kappa_{+}$. Note that while $\avg{\hat J_{c}(t)}$ is generally complex, it is real in the steady state.
	}
   \label{fig:chiJ}
\end{figure}

Using the result in \cref{eq:ctchi}, we find that the average value of this output quadrature is given by:
\begin{equation}
	\avg{\hat X_{\rm out}^{\ga}(t)} =  \sqrt{\frac{8 g^2}{\kappa_+}}  \Im\br{e^{i\ga}\avg{\hat J_{c}(t)}}.
\label{eq:Xval}
\end{equation}
In the steady state, our spin relaxes to a thermal state, and in general the value of  $\avg{\hat{J}_c(t)}$ will be non-zero and reflect
the overall size of the spin. It will thus be related to the initial mechanical phonon number, as illustrated in \cref{fig:chiJ}.  

One finds from \cref{eq:chi,eq:rhoJss} that in the steady-state $\avg{\hat J_{c}(t)}$ is purely real, implying that the optimal choice of quadrature corresponds to an angle $\alpha = \pi/2$.  This corresponds to measuring the phase quadrature of the output light leaving the primary mode.

The simplest way to extract the phonon number is to first turn on the measurement by driving the primary photonic mode, and then integrating
the output homodyne current (which is proportional to $\hat X_{\rm out}^{\ga}$) for a time $t$.  For sufficiently long integration times, the signal of phonon number in the average homodyne current will be resolvable above the noise.  For weak couplings,  the noise in the homodyne current is simply due to vacuum noise and is Gaussian.  As a result, standard calculations \cite{Clerk2010,Jayich2008} indicate that the mechanical Fock state $\hat{n}_b = n+1$ can be distinguished from   
Fock state $\hat{n}_b = n$ when the measurement time is sufficiently long, namely
\begin{equation}
	t > \gep^{-1 }\left(\tfrac{1}{j_{n}} \right)^2  \tau_{\rm meas}
\end{equation}
where \YYrev{$\gep$ is the measurement efficiency, including the effects of internal cavity loss,}
\begin{equation}
	\tau_{\rm meas} = \gk_{+}/g^{2}.
	\label{eq:tmeas}
\end{equation}
and the prefactor is determined by
\begin{equation}\begin{split}
j_n  =  4\Im\Big[e^{i\ga} \Big(& \avg{\hat J_{c}\p{t\to \infty}  }_{n_b = n+1} 
	\\ & - \avg{\hat J_{c}\p{t\to \infty}  }_{n_b = n}\Big)\Big].
\end{split}\end{equation}
As one can see from Fig.~\ref{fig:chiJ}, for a broad range of  $G / \delta \Omega$, the prefactor $1/j_n^2$ is of order unity, and the timescale of the measurement is set by $\tau_{\rm meas}$ as given in \cref{eq:tmeas}. 

We stress that the time scale required to distinguish between different mechanical Fock states in our setup is \emph{parametrically} shorter than that obtained in the proposal of Ludwig et al.~\cite{Ludwig2012}. Their scheme requires a measurement time which scales as $t \sim \p{\gd\gO/G}^{2}\tau_{\rm meas}$ to resolve two adjacent Fock states. However, the setup requires an interaction that is sufficiently non-resonant such that hybridization of phonons and photons is weak, and such that a perturbative treatment is applicable.  In practice, this requires both the conditions $g \ll |\gd \gO|$ and $G \ll |\gd \gO|$, yielding a measurement time which scales inversely with a small parameter. 
Our approach is not constrained in the same way, and can operate much closer to perfect resonance, where $G \sim |\gd \gO|$. As a result, for the same value of $g$, our approach yields greatly enhanced measurement speed. This in turn implies that an equally rapid measurement can be achieved with a smaller optomechanical coupling:  where Ludwig et al.~use $g = 3\gk_{+}$ to resolve phonon number states after $t = 50\gk_{+}^{-1}$ (see Ref.~\onlinecite{Ludwig2012} Fig.~3), \cref{eq:tmeas} implies our shelving protocol could do the same in a system where $g = 0.3\gk_{+}$.

\subsection{Optimal parameters\label{sec:optpar}} 

Our system has a fairly large number of tunable parameters. In particular, both non-zero components of the effective field $\vec B = (G,0 , \gd \gO)$ can be selected for optimal performance.  We remind the reader that $G$ is the many-photon coupling strength and $\gd \gO$ the mismatch between the mechanical frequency and photonic mode splitting. 

A key parameter to optimize the size of the signal in the long-time homodyne current which encodes the initial phonon number.  From \cref{eq:Xval}, this is directly determined by the steady-state value of $\avg{\hat J_{c}}$.  In steady state, we obtain from \cref{eq:chi,eq:rhoJss},
\begin{equation}
	\avg{\hat J_{c}}\to 
	\frac{B_{x}}{B}\avg{\avg{\hat J_{\parallel}}}_{T_{\rm eff}}
\label{eq:Jcss}
\end{equation}
where $\avg{\avg{\hat J_{\parallel}}}_{T_{\rm eff}}$ is the thermal average of $\hat J_{\parallel} = \vec{e}_B \cdot \vec{\hat{J}}$ taken at a temperature
$T_{\rm eff}$ given by \cref{eq:Teff}.

It follows from these expressions that to have an appreciable signature of the phonon number in the steady-state value of $\avg{\hat J_{c}}$ we need $B_x$ and $B_z$ to be comparable. A finite value of $B_x$ (i.e. many-photon coupling $G$) is needed to ensure that the steady state has a transverse component that can be detected by the primary photonic mode, as this mode couples to $\hat{J}_+$.  A non-zero value of $B_z$ (i.e.~mismatch between mechanical frequency and photonic mode splitting) is needed so that the effective temperature describing the steady-state is finite and the spin is not in a completely depolarized state.  

To make the above discussion more concrete, consider the simple case where we have one phonon intially, and our effective spin corresponds to a spin $1/2$.  There,
\begin{equation}
\avg{\hat J_{c}\p{t\to\infty}}_{n_{b} = 1}= \frac{B_{x}B_{z}}{B^{2} + B_{z}^{2}} =  \frac{G\gd\gO}{2G^{2} + \gd\gO^{2}}.
\end{equation}
\Cref{fig:chiJ} shows the steady state value of $\langle J_c \rangle$ for other values of initial phonon number.  We find the optimal ratio, for a system with one or a few initial phonons, is
\begin{equation}
	G/\gd\gO \approx 3/4.
\end{equation}

While the steady-state homodyne signal is not sensitive to the size of $\mathbf{B}$ (but rather only to the ratio $B_z / B_x$), the overall magnitude of the effective field is nonetheless important to our scheme.  If $B \gg \kappa_+$, then it follows from \cref{eq:Jmaster,eq:effrates} that the relaxation of the spin to its steady state is greatly suppressed.  In this limit, there is simply very little density of states in the primary $+$ mode for transitions that involve changing the energy of our spin.  We find that it is generally optimal to have the spin reach the steady state on a time scale much shorter than $\tau_{\rm meas}$, requiring $B \ll \kappa_+$. That is, one should avoid the  many-photon optomechanical strong coupling regime. 

Finally, to have the simple shelving dynamics depicted in Fig. 1b, the coherent oscillations between $c_{-}$ and b should be underdamped, leading to the requirement 
$(G \sim B) \gg g^2 / \kappa_+$. This condition is easy to meet in experiment given the typical weakness of single-photon optomechanical coupling strengths. Putting these conditions together, we find that the optimal regime for phonon number measurement requires:
\begin{equation}
	g^{2}/\kappa_+ \ll  \left( B  = \sqrt{G^2 + (\delta \Omega)^2 } \right) \ll \gk_{+}.
\end{equation}

An alternative approach to the measurement which results in faster measurement rates but is more complicated to implement experimentally is discussed in \cref{sec:Gmeas}.

\subsection{Resolving Fock states}

We next present numerical results for the measurement output of our scheme based on simulations of the  full master equation dynamics, including $\hat c_{+}$. We construct an estimator for phonon number from the integrated homodyne current:
\begin{equation}\begin{split}
	& \hat n_{b}^{\rm meas}\p{t} \equiv
	\\ & \qquad \br{ \sqrt{\tfrac{8 g^2}{\gk_{+}}}\tfrac{G \gd\gO}{2G^{2}+\gd\gO^{2}}}^{-1}
		\cdot \left( \frac{1}{t}
		\int_0^t d \tau 
			\hat X_{\rm out}^{\pi/2}(\tau) \right),
\label{eq:nmeas}
\end{split}\end{equation}
This estimator has been constructed so that at long times  
\begin{equation}
	\avg{\hat n_{b}^{\rm meas}(t)} \to \fopt{n_{b,{\rm init}} & n_{b,{\rm init}} = 0, 1 \\ f[n_{b,{\rm init}}] \, n_{b,{\rm init}} & n_{b,{\rm init}} \ge 2}
\end{equation}
where $n_{b,{\rm init}}$ is the initial phonon number, and $f[n]$ can be calculated from the effective temperature above, and goes to 
$\br{\tfrac{2G \gd\gO}{2G^{2}+\gd\gO^{2}}}^{-1}$ for $n \to\infty$.

\begin{figure}[t] 
   \centering
   \includegraphics[width=\columnwidth]{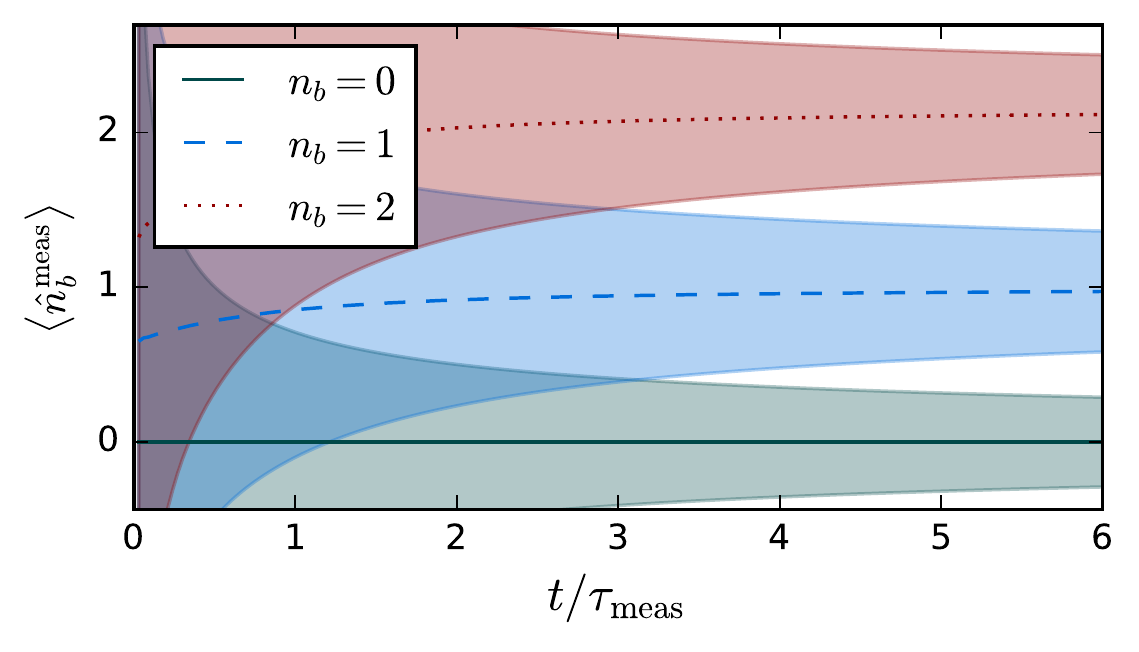} 
   \caption{Behavior of the estimated photon number $n_{b}^{\rm meas}(t)$ as a function of integration time $t$. The estimator, obtained from the time-integrated homodyne current, is described in \cref{eq:nmeas}, and time is given in units of $\tau_{\rm meas} = \gk_{+}/g^{2}$.
   Curves correspond to the mean value of the estimator, shaded regions correspond to the standard deviation.  Different curves correspond
   to different initial mechanical phonon numbers.  One sees that different initial mechanical phonon numbers become distinguishable
   on a time scale proportional to $\tau_{\rm meas}$.  
   Parameters correspond to $g = 0.01\gk_{+}, G = 0.1 \gk_{+}, \gd\gO = 0.13 \gk_{+}$, and to the ideal case where $\gamma = \kappa_{-}  =0$.
}
   \label{fig:QND}
\end{figure}

The behavior of the measurement output, expressed in terms of the phonon number estimator $\hat n_{b}^{\rm meas}(t)$, are presented in \cref{fig:QND}, again for the ideal case of $\gamma = \kappa_{-} = 0$.  As expected from our analysis, different phonon numbers are clearly resolvable after an integration time on the order of $ \tau_{\rm meas}$. The details of these numerical calculations are given in \cref{sec:numerics}.

Note that in this ideal situation where there is no spurious dissipation, the resolving power of our measurement continues to increase indefinitely as the integration time $t$ is increased.  This reflects the shelving nature of the dynamics.  We now go on to address the additional limitations placed by unwanted dissipation.

\begin{figure*}[t] 
   \centering
   \includegraphics[width=2\columnwidth]{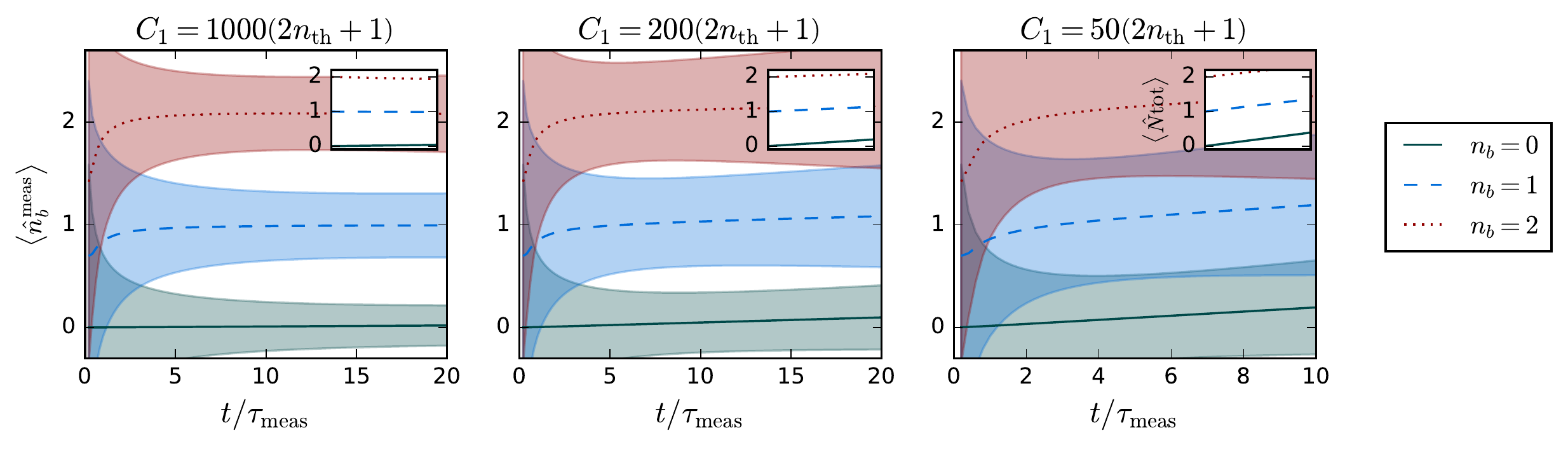} 
   \caption{Behavior of the estimated photon number $n_{b}^{\rm meas}(t)$ as a function of integration time $t$ when spurious dissipation is taken into account. As in \cref{fig:QND}, the estimator is described in \cref{eq:nmeas}, and time is in units of $\tau_{\rm meas} = \gk_{+}/g^{2}$. 
      Curves correspond to the mean value of the estimator and shaded regions correspond to the standard deviation. Insets show the behavior of $\avg{\hat N_{\rm tot}}$ over the same range of times.
   We consider parameters where mechanical dissipation is the key limit to the scheme; as such, the relevant parameter is the single-photon cooperativity $C_{1} = 4g^{2}/\gk_{+}\gamma$.
   When spurious dissipation is negligible (left), $\avg{\hat N_{\rm tot}}$ is nearly conserved and different initial mechanical phonon numbers $n_{b}$ are easily resolved if we integrate on a time scale longer than $\tau_{\rm meas}$. When spurious dissipation is relevant but sufficiently small, the integrated homodyne current can still be used to determine the initial phonon number $n_{b}$ at intermediate times. At longer times, the mechanical system begins to relax into a thermal equilibrium independent of its initial state. 
   If the rate of spurious dissipation is too fast (right) the system relaxes before differences in $n_{b}$ are resolvable.  
   Parameters used are $g = 0.1\gk_{+}$, $G = 0.1\gk_{+}$, $\gd\gO = 0.13\gk_{+}$, $\gk_{-} = 10^{-4}\gk_{+}$, $n_{\rm th} = 100$ and $\gamma$ varies. The details of the numerical integration are given in \cref{sec:numerics}.
   }
   \label{fig:gamma}
\end{figure*}


\section{Spurious dissipation: mechanical and auxiliary mode losses\label{sec:gamma}}

Our analysis so far has treated the ideal case, where the only dissipation is the necessary coupling between the primary photonic mode and the waveguide (or transmission line) used to collect the output field which serves as the measurement record. This idealized analysis gives a useful picture for understanding the more realistic experimental case where there is also mechanical dissipation (damping rate $\gamma$) and damping of the auxiliary mode (damping rate $\gk_-$). 

All three forms of spurious dissipation (mechanical loss, mechanical heating, auxiliary mode loss) cause the quantity $\hat{N}_{\rm tot}$ to change, disrupting the shelving physics that is at the heart of our scheme.  If spurious dissipation changes $\hat{N}_{\rm tot}$, the simple correspondence between the measurement record and the initial mechanical phonon number is lost.  
Focusing on small initial phonon numbers, the fastest rate at which such spurious processes occur is given by $\max\br{\gk_{-},\gamma \p{2n_{\rm th}+1}}$.  Thus, one requires that the measurement essentially occur on a timescale much shorter than the inverse of this rate.  This translates to the condition
\begin{equation}
	\left(\tau_{\rm meas}^{-1} = \frac{g^{2}}{\gk_{+}} \right) \gg \max\br{\gk_{-}, \gamma\p{2n_{\rm th} + 1}}.
	\label{eq:QNDCondition}
\end{equation}
In \cref{fig:gamma} we show the results of detailed master equation simulations including all forms of spurious dissipation which show this intuitive understand is correct.

It is interesting to consider two relevant limits of the condition in Eq.~(\ref{eq:QNDCondition}).  
In the case where $\kappa_{-}$ is much faster than the mechanical heating $\gamma (2 n_{\rm th}+1)$, the condition reduces to 
$g^{2} \gg \gk_{+}\gk_{-}$.  This is the same constraint that limits other proposals for measuring phonon number in two-cavity optomechanical systems \cite{Miao2009,Ludwig2012}.  

\YYrev{While it is beyond current experimental capacities,} given recent progress in constructing two-mode optomechanical systems with extremely low auxiliary-cavity dissipation \cite{Paraiso2015}, it is worth considering the opposite limit, where  $\gamma \p{2 n_{\rm th} + 1}$ dominates.  In this case, our condition reduces to 
\begin{equation}
	\left( C_{1} = \frac{4g^{2}}{\gk_{+} \gamma} \right) \gg 2n_{\rm th} + 1.
	\label{eq:QNDreq}
\end{equation}
where we have introduced the single-photon cooperativity $C_1$.  This regime is illustrated in \cref{fig:gamma}. Numerically, we find the requirement is ${C_{1} \gtrsim 100\p{2n_{\rm th}+ 1}}$ to distinguish between the vacuum state and a single phonon, if the mechanics are the dominant source of spurious dissipation.

\YYrev{Our work represents one of the only protocols or effects where this parameter plays a crucial role. By comparison, in Ref.~\onlinecite{Ludwig2012}, even in the limit $\gk_{-} = n_{th} = 0$, the QND condition reduces to  $C_{1}\tfrac{G^{2}}{\gd\gO^{2}}\gg 1$. As their scheme relies on the assumption $G\ll \gd\gO$, this requires $C_{1}$ to be parametrically larger than \cref{eq:QNDreq} does.}

\section{Outlook}

We have presented and analyzed  a new approach for QND phonon number measurement in quantum optomechanics, one which adapts the successful shelving strategy used in trapped-ion systems.  
By considering regimes where the driving of the photonic system is strong (and not perturbative), we are able to obtain measurement rates that are parametrically larger than previously considered approaches.  At the level of theory, our approach of mapping a bosonic problem onto a spin system could be useful in more complex optomechanical systems where there are again conserved quantities akin to the excitation number $\hat N_{\rm tot}$ in our system.  

Our approach is most suited  to systems where the auxiliary (unmeasured) photonic mode has an extremely low dissipation rate; recent experiments on two-cavity microwave-frequency optomechanical systems \cite{Toth2016} present a promising route to this regime. \YYrev{The limiting factor remains the strong coupling requirement, ${g^{2}/\gk_{+} \gg \max\br{\gk_{-},\gamma\p{2n_{th}+1}}}$ (see \cref{eq:QNDCondition}). Current experiments in microwave cavity systems have been in the range ${g/2\pi \sim 10^{2} \unit{Hz}}$, ${\gk_{\pm}/2\pi \sim 10^{5}-10^{9}\unit{Hz}}$, but recent proposals have been made for fabrication of microwave devices where both would be in the MHz range \cite{Paraiso2015}.}

\section*{Acknowledgements}

This research was undertaken in part thanks to funding from NSERC and the Canada Research Chairs program.


\appendix

\newpage
\section{Analogy with electron shelving\label{sec:shelvinganalogy}}

\begin{figure}[htb] 
   \centering
   \subfloat[Shelving in ions \label{fig:analogy-ions}]{\includegraphics[width=0.47\columnwidth]{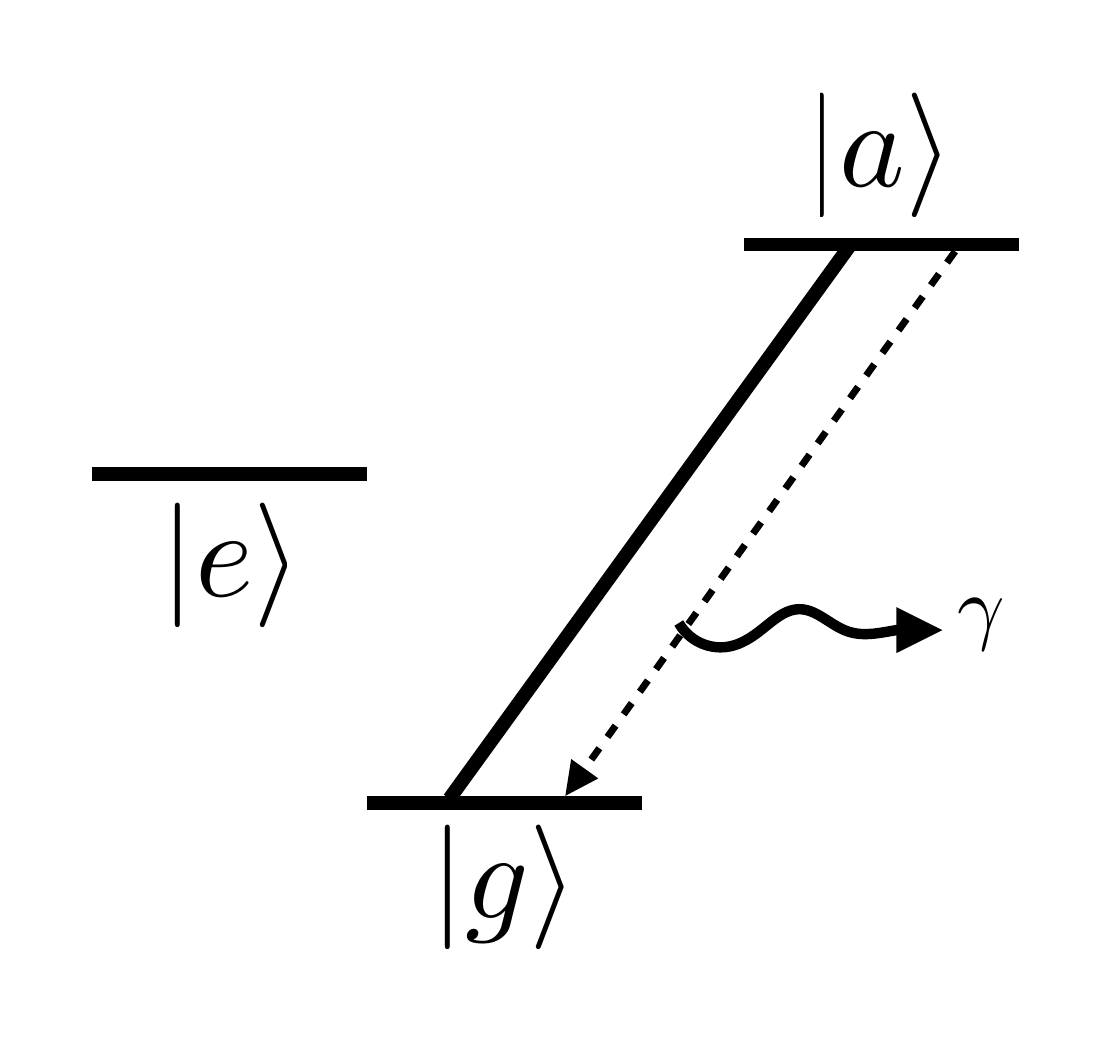}}\hfill
   \subfloat[Shelving in optomechanics]{\includegraphics[width=0.47\columnwidth]{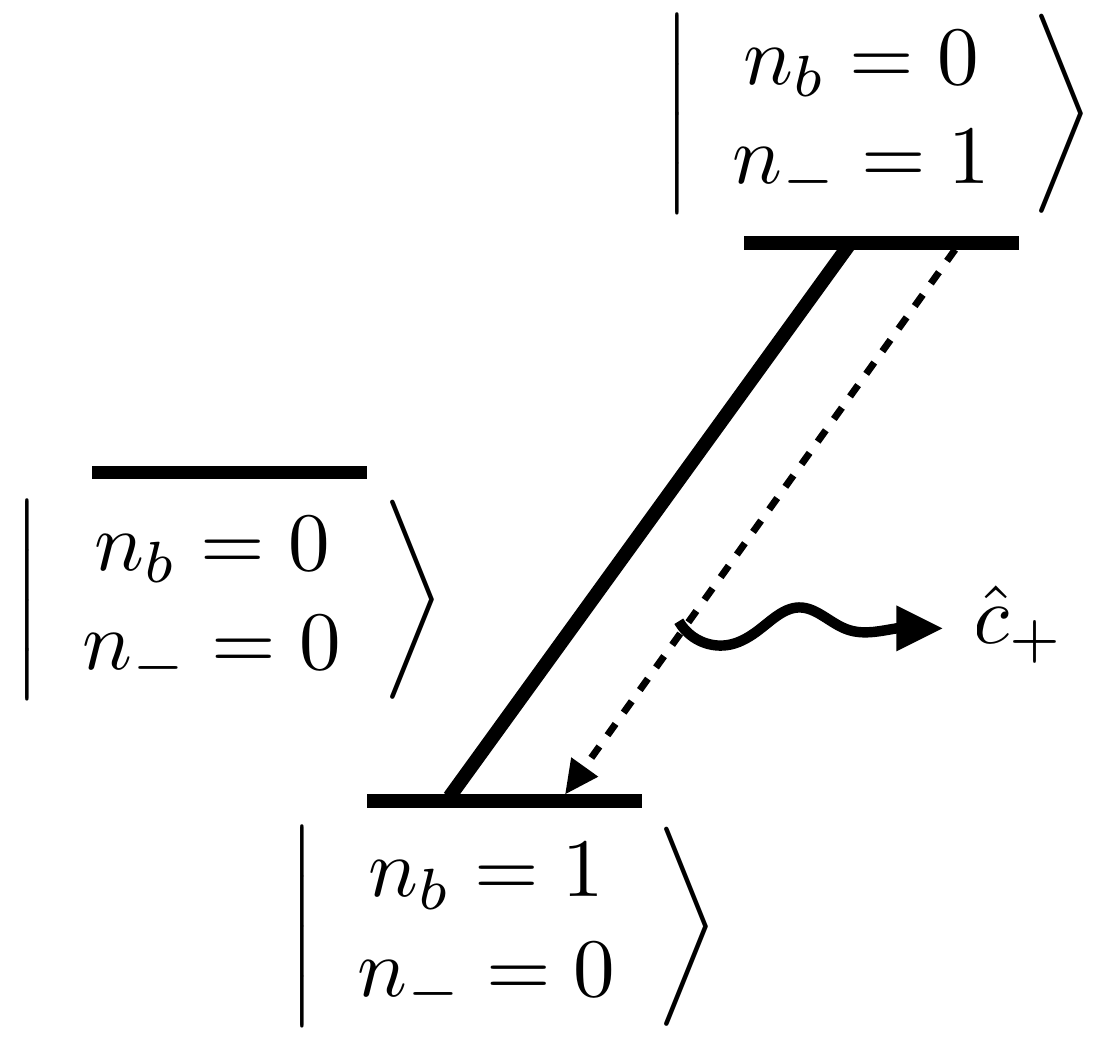}\label{fig:analogy-om}}

   \caption{The electronic shelving protocol, in ions \cite{Javanainen1986} and in our optomechanical system.  {\bf (a)} In Ref.~\onlinecite{Nagourney1986}, when dealing with laser cooled $\rm Ba^{+}$, the long-lived states are $\ket{g} = {\rm6^{2}S_{\half}}$ and $\ket{e} = {\rm5^{2}D_{\half[5]}}$, while the auxiliary modes are the short-lived $\ket{a} = {\rm6^{2}P_{\half[5]}}, {\rm5^{2}D_{\half[3]}}$. The thick line represents the driving lasers while the dashed line represents decay with photoemission. 
   {\bf (b)} The same diagram represents the optomechanical system for $\hat N_{\rm tot} = 0,1$. Here, the thick lines represents the driving $G$ while the dashed line represents decay with emission of a primary photon.
   }
   \label{fig:analogy}
\end{figure}

The shelving protocol, pioneered by Hans Dehmelt \cite{Dehmelt1968,Nagourney1986,Javanainen1986}, is a form of QND measurement of the state of two-level atomic system. Given an ion in its ground state ($\ket{g}$) or a metastable excited state ($\ket{e}$), the protocol makes use of a short-lived auxiliary state ($\ket{a}$). By applying a resonant laser drive we induce Rabi oscillations between $\ket{g}$ and $\ket{a}$; this is accompanied by the spontaneous decay process $\ket{a}\to\ket{g}$. The detection of photoemission in the appropriate wavelength is then confirmation that the atom is not in the excited state  $\ket{e}$. However, throughout the process, the atom cannot be said to be in the original ground state $\ket{g}$. Instead it is in some hybrid state of $\ket{g}$ and $\ket{a}$. This process is sketched out in figure \cref{fig:analogy-ions}.

Notice that the rate of spontaneous emission is low, and distinguishing the signal from the noise requires a prolonged measurement similar to what is discussed in \cref{sec:meas}. Thus the original shelving protocol also constitutes a weak measurement.

The analogy to our measurement scheme is most direct in the case where we are trying to distinguish an initial mechanical phonon number of $\hat n_{b} = 0$ or $\hat n_{b} = 1$. Here, the system is originally in the state $\ket{1,0}$ or $\ket{0,0}$ (where $\ket{n_{b},\hat n_{-}}$ denotes a Fock state with the respective number of phonons and auxiliary photons). The application of the drive $G$ generates Rabi oscillations between $\ket{1,0}$ and $\ket{0,1}$, while a decay process allows the relaxation $\ket{0,1}\to\ket{1,0}$ with emission of a primary photon, first into the cavity and then out of it through the dissipation channel $\gk_{+}$.   By detecting these emitted photons, we can determine the initial phonon number.  This process is sketched out in figure \cref{fig:analogy-om}.

Our scheme takes the basic shelving concept further, allowing one to distinguish between multiple possible values of $\hat{n}_b$ while remaining non-destructive.  As discussed in the text, information on the phonon number is encoded in the rate at which emitted photons are produced.  

\section{Optimized measurement prototocol}
\label{sec:Gmeas}

\begin{figure}[t] 
   \centering
   \includegraphics[width=\columnwidth]{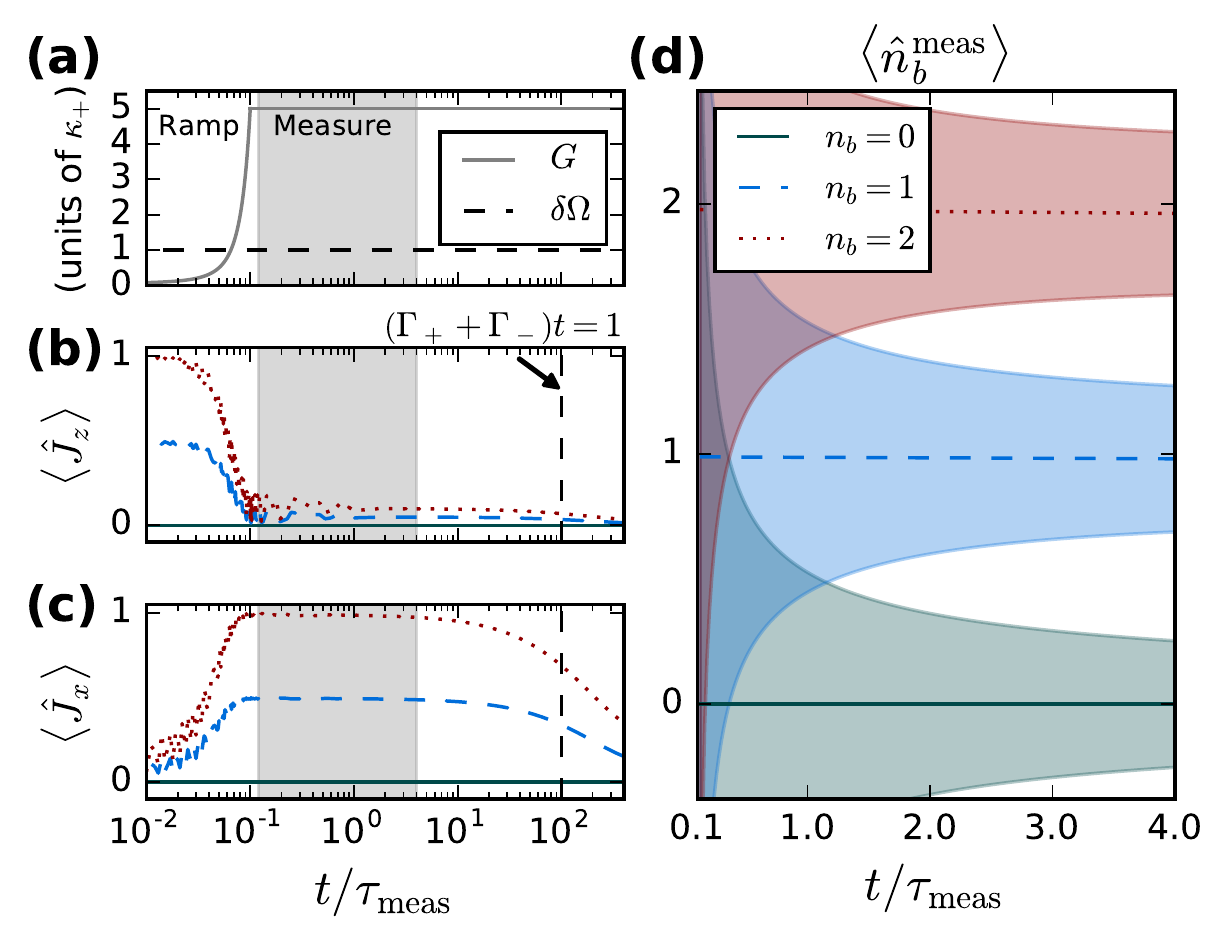} 

   \caption{QND mechanical phonon number measurement using the alternate measurement protocol discussed in \cref{sec:Gmeas}.  
   {\bf (a)} The measurement protocol. At time $t = 0$ the system is initiated in a pure state. From time $0<t<\p{t_{f} = 0.1\tau_{\rm meas}}$, the driving is ramped up, with $G\p{t} = G_{f}/100^{1-t/t_{f}}$. We then keep $G$ fixed and the measurement protocol outlined in \cref{sec:meas} is performed. 
   {\bf (b-c)}~Behavior of the spin system. Initially, with $\gd\gO \gg G$, the spin is aligned along $\mathbf{e}_{z}$. When the driving is ramped up semi-adiabatically, the direction of the spin rotates with it to point along $\mathbf{e}_{z}$. At a later time, $t\sim \p{\Gamma_{+} + \Gamma_{-}}^{-1}$ (see \cref{eq:effrates}), the magnitude of the spin decays into its steady state value.
   {\bf (d)} The phonon number estimator as a function of time during the measurement. This plot corresponds to the gray shaded area on the left. Curves correspond to the mean value of the estimator, and shaded regions correspond to the standard deviation. The different phonon numbers can be differentiated about three times faster than in \cref{fig:QND}.  
   The time is in units of $\tau_{\rm meas} = \gk_{+}/g^{2}$. The parameters used here are $g = 0.01\gk_{+}$, $\gd\gO = \gk_{+}$, $G_{f} = 5\gk_{+}$ and with no spurious dissipation, $\gk_{-} = \gamma = 0$. Details about the numerical calculation are given in \cref{sec:numerics}.
   }
   \label{fig:QND-alt}
\end{figure}

As discussed in \cref{sec:optpar}, phonon number measurement in the steady state is subject to two conflicting constraints: the magnitude of $\avg{\vec{\hat J}}$ is proportional to $B_{z}/B$ while the measurable cavity field is coupled to the transverse components,  $\hat c_{+} \sim \avg{\hat J_{+}} \sim B_{x}/B\abs{\avg{\vec{\hat J}}}$.

These competing requirements on the effective field can be avoided by arranging parameters so that the system has a long relaxation time, and performing the measurement in the transient period before a steady state is reached. We can then align the spin entirely along the axis of measurement and increase the magnitude of the cavity field. As we have seen in \cref{eq:effrates}, a long relaxation time is achieved in the regime $B\gg \gk_{+}$. We therefore suggest the following protocol:
\begin{enumerate}
\item The system is set up with some large energy mismatch $\gd\gO \gtrsim \gk_{+}$. This ensures a slow relaxation to the steady state once the primary mode drive is turned on,  the relevant rates being ${\Gamma_{\pm} \sim \p{\gk_{+}/B}^{2}\tau_{m}^{-1}}$.
\item With the driving turned off, $G = 0$, the mechanics are prepared in their initial state. As discussed in \cref{sec:model}, the auxiliary mode remains at zero population in the absence of driving. This corresponds to the spin pointing along $\mathbf{e}_{z}$.
\item We then ramp up the primary mode drive semi-adiabatically, until $G\gg \gd\gO$. In the spin language, we increase the $B_{x}$ component until the effective field is nearly parallel with $\mathbf{e}_{x}$.
\item If the driving is increased sufficiently slowly, the spin itself will track the direction of the field and finally point along $\mathbf{e}_{x}$. We make use of the separation of scales to ramp over a time $t_{\rm ramp}$ such that
\begin{equation}
B^{-1}\ll \gk_{+}^{-1} \ll t_{\rm ramp}\ll \tau_{\rm meas}.
\end{equation}
\item Finally, we perform a homodyne measurement as described in \cref{sec:meas}. Here we use the estimator 
\begin{equation}
 \hat n_{b}^{\rm meas}\p{t} \equiv
	\br{\sqrt{\tfrac{2g^{2}}{\gk}}}^{-1} \cdot\left( \frac{1}{t}\int_0^t d \tau \hat X_{\rm out}^{\pi/2}(\tau) \right).
\label{eq:nestalt}
\end{equation}
\end{enumerate}

This procedure is illustrated in \cref{fig:QND-alt}. We see that the required measurement time is about three times shorter for this set of parameters.

\section{Details of Numerical Simulations\label{sec:numerics}}
The numerical calculations presented throughout this work are the results of full master equation treating each of the three bosonic modes in our system:
\begin{equation}\begin{split}
\dot{\hat \rho} = i\br{\hat\rho, \hat H_{\rm eff}} & + \gk_{+}\mathcal D\br{\hat c_{+}}\cdot\hat \rho + \gk_{-}\mathcal D\br{\hat c_{-}}\cdot\hat \rho
\\ &  + \br{\gamma\p{n_{\rm th} + 1}\mathcal D[\hat b] + \gamma n_{\rm th}\mathcal D[\hat b\dg]}\cdot\hat \rho,
\label{eq:masternum}
\end{split}\end{equation}
where $\hat H_{\rm eff}$ is given in \cref{eq:Heff} and 
\begin{equation}
\mathcal D\br{\hat a}\cdot{\hat \rho} = \hat a\hat \rho \hat a\dg - \half\p{\hat a\dg\hat a \hat\rho + \hat\rho \hat a\dg\hat a}
\end{equation}
is the standard Lindblad superoperator. 

The numerical simulations were performed using QuTiP \cite{Johansson2013}. We truncated the Hilbert space to ${\hat c_{+}\dg\hat c_{+}  = 0,1,2}$, and to $0\le\hat N_{\rm tot} \le 6$ for simulations involving spurious dissipation (truncation is not needed if $\gk_{-} = \gamma = 0$). We then solved the eigenvalue problem defined by \cref{eq:masternum} and calculated $\hat\rho\p{t}$ at any time as a sum of exponentials.

The initial state in all cases was $\hat \rho\p{0} = \ket{\psi_{0}}\bra{\psi_{0}}$ where
\begin{equation}
\hat b\dg\hat b\ket{\psi_{0}} = n_{b}\ket{\psi_{0}} \qquad \hat c_{\pm}\ket{\psi_{0}}  = 0.
\end{equation}

To simulate the driving of the primary mode, the parameter $G$ is initially ramped from zero to its final value. For \cref{fig:QND,fig:gamma} we ramp $G$ linearly over a time span of $\gk_{+}^{-1}$, while in \cref{fig:QND-alt} we increase it exponentially as explained in the figure caption. Experimentally these behaviors can be achieved using pulse-shaping techniques.

The standard deviations plotted are given by
\begin{equation}
\Delta n_{b}^{\rm meas}\p{t} = \sqrt{\avg{\p{\hat n_{b}^{\rm meas}\p{t}}^{2}} - \avg{\hat n_{b}^{\rm meas}\p{t}}^{2}}.
\end{equation}
They were calculated using the quantum regression theorem \cite{Gardiner2004}, taking into account the full dynamics of the system and not just cavity vacuum noise.

\bibliography{ResonantQND-rev1.bbl}

\end{document}